\documentclass[preprint,prd,amsmath,amssymb,nofootinbib,tightenlines,floatfix,superscriptaddress]{revtex4}

\usepackage{colordvi}
\usepackage[dvipsnames,usenames,pdftex]{color}
\usepackage[pdftex]{graphicx,epsfig}
\usepackage{rotating}
\usepackage{wrapfig}
\usepackage{epsf}
\usepackage{hyperref}
\usepackage{fancybox}
\usepackage{amsmath,amsthm,amssymb,amsfonts}
\usepackage{epsfig}
\usepackage{latexsym}
\usepackage{siunitx}
\usepackage{subfig}
\usepackage{graphicx}

\begin{document}

%\title{Community input to the European Strategy on particle physics:\\ Searches for Permanent Electric Dipole Moments}
%\author{{\bf edited by:}~\\ M.~Athanasakis-Kaklamanakis, M.~Au, R.~Berger, S.~Degenkolb, J.~De Vries, S.~Hoekstra, A.~Keshavarzi, N.~Neri, D.~Ries, P.~Schmidt-Wellenburg, and M.~Tarbutt,\\~\\
%{\bf endorsed by the\\
%European EDM projects and collaborations:}\\
%G.~{Carugno -- DOCET EDM experiment}\\
%P.~{Fierlinger -- PanEDM collaboration }\\
%B.~{Filippone -- nEDMsf collaboration}\\
%S.~Hoekstra -- NL-eEDM collaboration\\
%B.~{Lauss  -- nEDM at PSI collaboration}\\
%N.~{Neri -- Aladdin collaboration}\\
%F.~{Piegsa -- Beam EDM collaboration}\\
%Y.~K.~{Semertizidis -- pEDM collaboration}\\~\\
%{\bf individually endorsed by:}\\
%A.~{Borschevsky -- University of Groningen, The Netherlands}\\
%U.~{van Kolck -- ECTStar, Trento, Italia}}

%\begin{abstract}

%\end{abstract}

%\author{edited by: M.~Athanasakis-Kaklamanakis}
%\affiliation{Imperial College London, UK}
%\author{M.~Au}
%\affiliation{CERN}
%\author{D.~Ries}
%\affiliation{Paul Scherrer Institute, CH}

%\collaboration{endorsed by the\\
%European EDM projects and collaborations:}

%\collaboration{col2}

%\maketitle

\begin{titlepage}
\begin{center}
    \textbf{\large{Community input to the European Strategy on particle physics:\\\vspace{0.1cm} Searches for Permanent Electric Dipole Moments}}
\end{center}
\begin{center}
\textbf{edited by:}

M.~Athanasakis-Kaklamanakis (\textit{Imperial College London, United Kingdom}), M.~Au (\textit{CERN, Geneva, Switzerland}), R.~Berger (\textit{University of Marburg, Germany}), S.~Degenkolb (\textit{Heidelberg University, Germany}), J.~De~Vries (\textit{University of Amsterdam, The Netherlands})
S.~Hoekstra (\textit{University of Groningen, The Netherlands}), A.~Keshavarzi (\textit{University of Manchester, United Kingdom}), N.~Neri (\textit{University of Milan and INFN Milan, Italy}), D.~Ries (\textit{Paul Scherrer Institute, Villigen, Switzerland}), P.~Schmidt-Wellenburg (\textit{Paul Scherrer Institute, Villigen, Switzerland}), and M.~Tarbutt (\textit{Imperial College London, United Kingdom})
\end{center}
\vspace{-0.3cm}
\begin{center}
\textbf{endorsed by the\\
European EDM  projects and collaborations:}
\vspace{0.1cm}

\textbf{Beam EDM collaboration} -- F.~Piegsa (\textit{U.~Bern, Switzerland})\\
\textbf{quMercury experiment} -- S. Stellmer (U.~Bonn, Germany)\\
\textbf{pEDM collaboration} -- Y.~K.~Semertizidis (\textit{KAIST, Daejeon, Korea})\\
\textbf{NL-eEDM collaboration} -- S.~Hoekstra (\textit{U.~Groningen, The Netherlands})\\
\textbf{PanEDM collaboration} -- S. Degenkolb (\textit{U.~Heidelberg, Germany}) 
and P.~Fierlinger (\textit{Technical University of Munich, Germany})\\
\textbf{HeXe collaboration} -- U.~Schmidt (\textit{U.~Heidelberg, Germany})\\
\textbf{Imperial eEDM collaboration} -- M.~R.~Tarbutt (\textit{Imperial College London, UK})\\
\textbf{ALADDIN collaboration} -- N.~Neri (\textit{University of Milan and INFN Milan, Italy}) and F.~Martinez-Vidal  (\textit{U.~Valencia and IFIC, Spain})\\
\textbf{EDMMA collaboration} -- D.~Comparat(\textit{LAC Orsay, France})
\textbf{DOCET EDM experiment} -- G.~Carugno (\textit{INFN Padua, Italy})\\

\textbf{muEDM collaboration} -- A.~Papa (\textit{INFN Pisa, Italy})\\
\textbf{nEDMSF collaboration} -- W.C.~Griffith (\textit{U. Sussex, UK}) and\\ M.~Jentschel (\textit{ILL, France})\\
\textbf{nEDM at PSI collaboration} -- B.~Lauss (\textit{Paul Scherrer Institute, Villigen, Switzerland})\\
\end{center}
\vspace{-0.3cm}
\begin{center}
\textbf{individually endorsed by:}\\
A.~Borschevsky (\textit{U.~Groningen, The Netherlands}), 
V.~Cirigliano (\textit{U.~Washington, USA}), 
J.~Dobaczewski (\textit{U.~Warsaw, Poland}), 
K.~Flanagan (\textit{U.~Manchester, UK}),
T.~Fleig (\textit{U.~Toulouse, France}),
M.~Kortelainen (\textit{U.~Jyväskylä, Finland}),
L.~Di~Luzio (\textit{INFN Padova, Italy}),
G.~Neyens (\textit{KU Leuven, Belgium}),
L.~Nies (\textit{CERN}),
G.~Onderwater (\textit{U.~Maastricht, The Netherlands}),
U.~van~Kolck (\textit{ECTStar, Trento, Italy}),
V.~Sanz (\textit{U.~Valencia, Spain}),
P.~Stoffer (\textit{PSI, Switzerland}),
O.~Vives (\textit{IFIC, Spain})
\end{center}
\vspace{-0.2cm}

\begin{center}
{\large Abstract}

Searches for electric dipole moments~(EDMs) in fundamental particles and quantum systems with spin are pivotal experiments at the intersection of low-energy and high-precision particle physics. 
These investigations offer a complementary pathway to uncovering new physics beyond the Standard Model, parallel to high-energy collider searches. 
EDM experiments are among the most sensitive probes for detecting non-standard time-reversal~(T) symmetry violations and, via the CPT theorem, CP-violation~(CPV). 
Current EDM measurements test new physics at mass scales in or above the $10-100\,$TeV range.\\
This community input to the European Particle Physics Strategy Update highlights the status of the field, and describes challenges and opportunities in Europe. 
\end{center}

\setcounter{page}{1}
\clearpage
\end{titlepage}

\hspace{1ex}

\section{Probing beyond Standard Model physics by searching for permanent electric dipole moments }
% Introduction one page relevance of the EDM searcher and their complementarity to HEP
% update old text:

Searches for electric dipole moments (EDM) of fundamental particles
and systems with spin (like neutrons, atoms, molecules, protons,
deuterons, muons, ...) are considered to be among the most
important particle physics experiments at the low energy, high
precision frontier: see e.g.~\cite{Raidal08, Chupp19, Cairncross19}. They provide an
alternative route to new physics, which is complementary to searches
at high energy colliders. EDM searches provide the most sensitive
tests of non-standard time reversal invariance violation and, by the
CPT-theorem, of CP-violation~(CPV). Generically, assuming couplings with a
CPV phase of order one, dimensional estimates for fundamental fermion EDMs today indicate new physics
should lie above mass scales of 10-100\,TeV~\cite{Engel:2013lsa, Chupp19}. Model-independent constraints, drawn from the full portfolio of measured EDM systems and interpreted within Standard Model Effective Field Theory, are presently limited by Standard Model uncertainties and therefore a target for significant improvements.

\begin{figure}
\centering
\subfloat[]{
  \includegraphics[width=.32\linewidth]{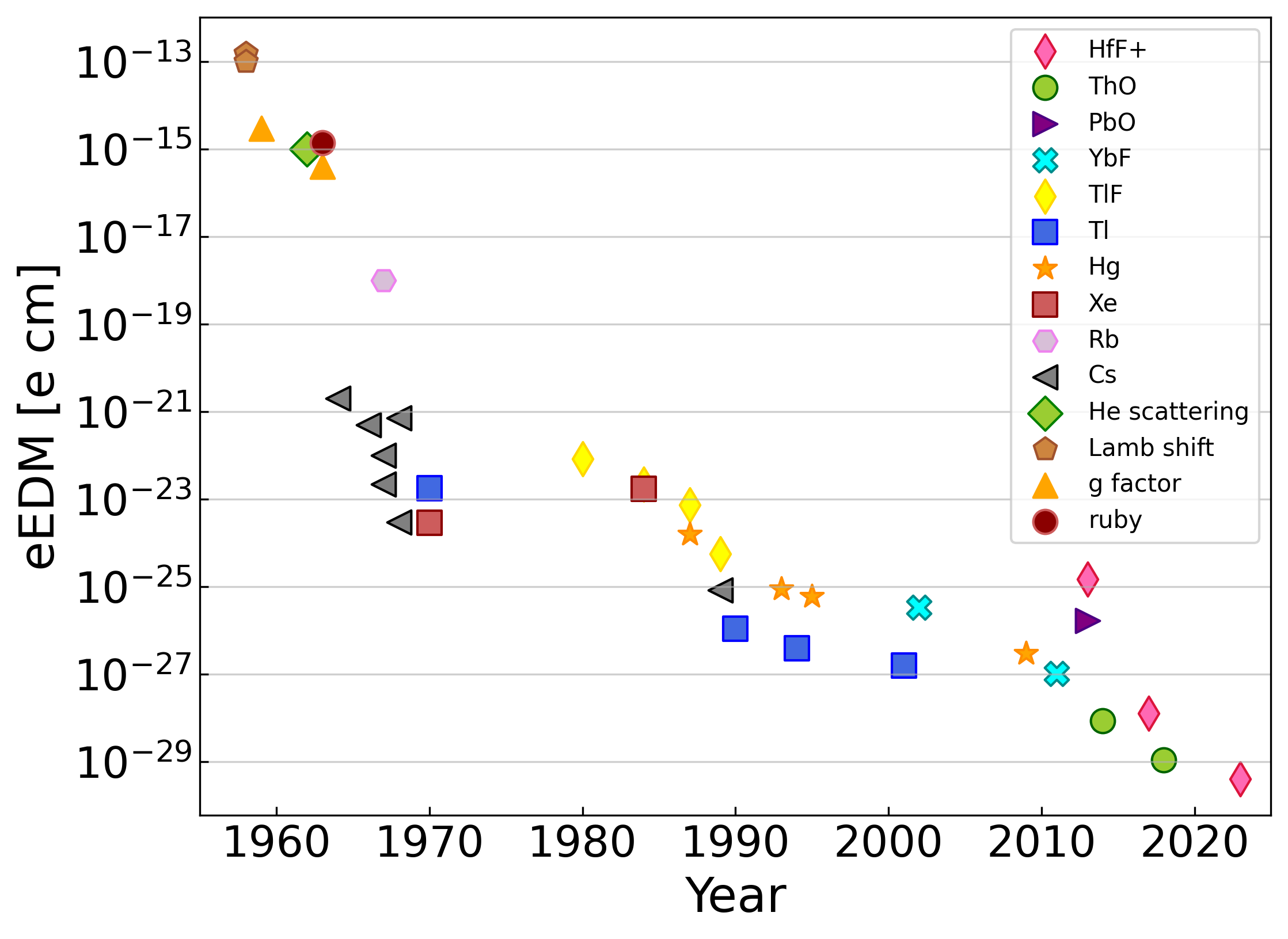}}
  \label{fig:sub1}
\hfill
\subfloat[]{
  \includegraphics[width=.32\linewidth]{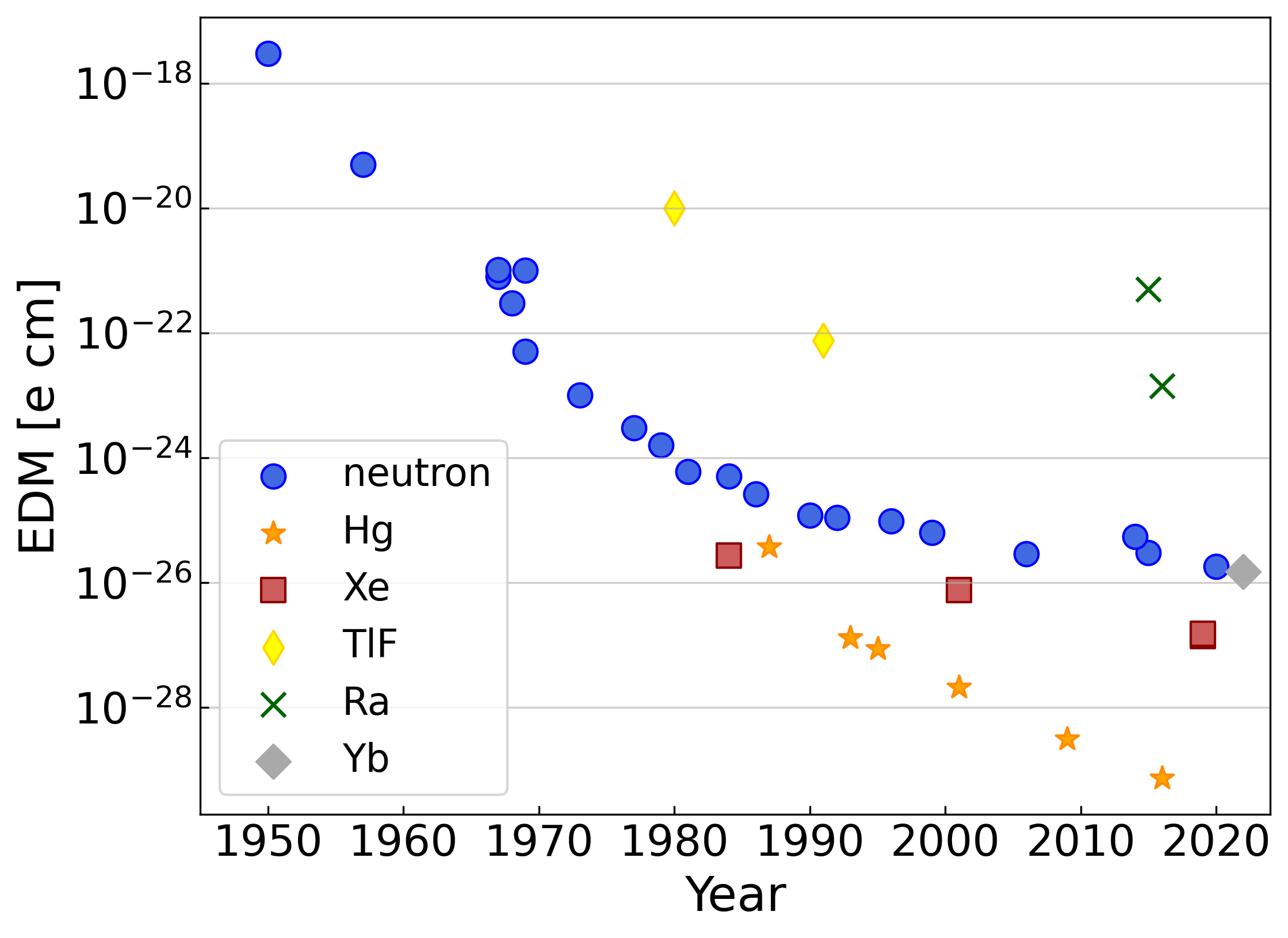}}
  \label{fig:sub2}
\hfill
\subfloat[]{
  \includegraphics[width=.32\linewidth]{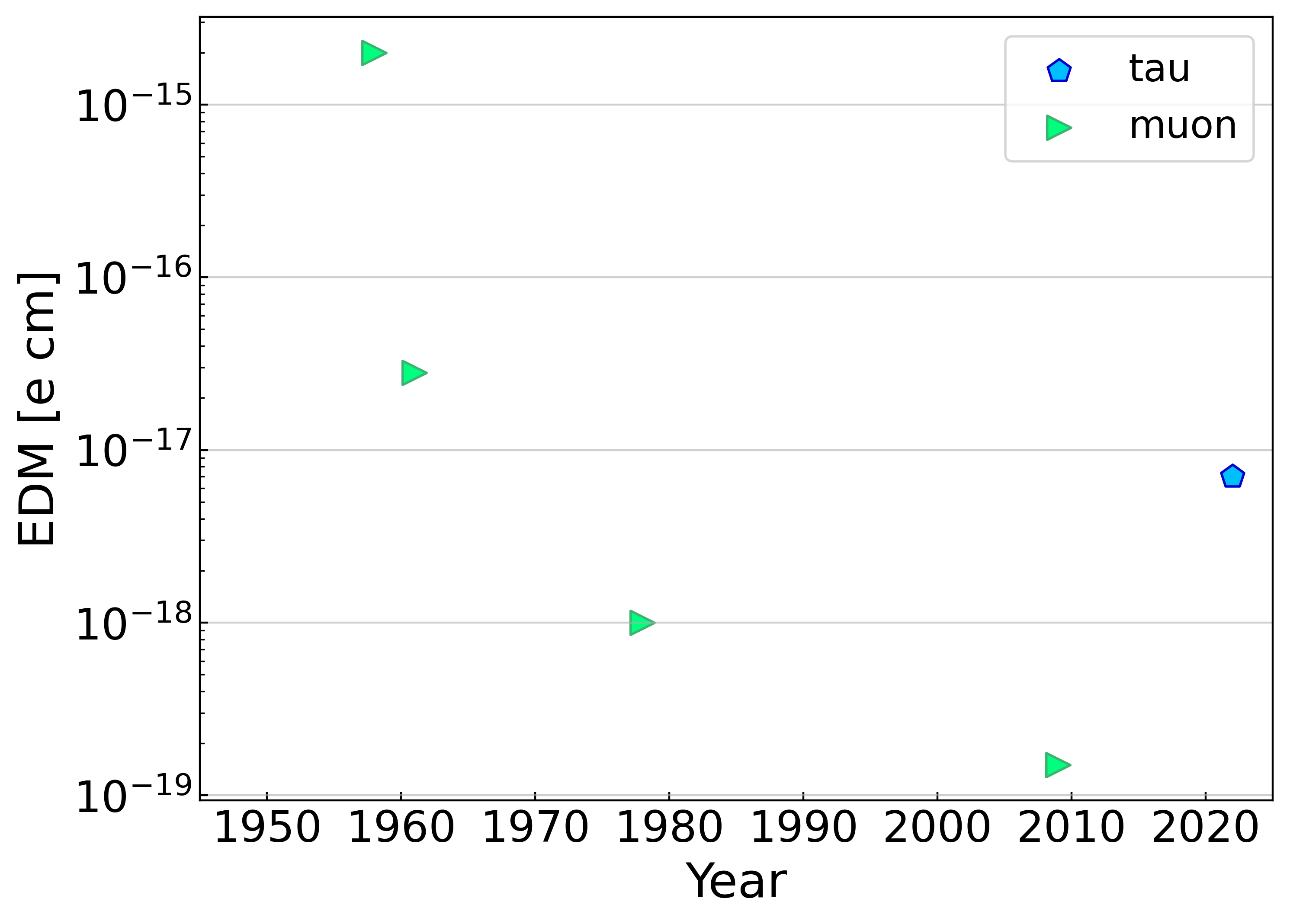}}
  \label{fig:sub3}
\caption{Progress of EDM measurement sensitivity over time, shown as a compilation of EDM limits measured in a selection of systems~\cite{Jayich_github}. (a)~Electron EDM limits from several systems, assuming the electron EDM as the only source of CP violation. (b)~Closed shell EDM limits assuming no specific source of CP violation. (c)~EDM limits of the muon and the tau, from direct measurements~\cite{Ben09,Belle2022}.}
\label{fig:test}
\end{figure}

Most EDM experiments search for an electric-field-induced spin precession of a neutral system, e.g.\ neutrons, atoms, or molecules, while meticulously controlling any magnetic field.
By reversing the electric field direction, the spin precession induced by the magnetic field cancels, making a possible EDM spin precession detectable. 
In most cases, where Ramsey's technique of separated oscillating fields~\cite{Ramsey1950} is used to measure the spin precession, the sensitivity scales with the inverse of the observation time $T$, the electric field $E$, and the square root of the number of observed particles $\sqrt{N}$. Table~\ref{tab:my_label} lists the most stringent EDM limits set to date, Fig.~\ref{fig:test} shows the historical progress of EDM searches, while Tab.~\ref{tab:EDMProspects} list planned and ongoing EDM searches and their sensitivity goals.

The observation of any new CPV physics would be a very
significant discovery. Interpretation of the experimental results
requires theoretical treatment on various levels. For instance, the
most sensitive limits on the EDM of the electron come from
experiments with charged and neutral molecules~\cite{ACME18,Roussy23}. These are sensitive both to the intrinsic electron EDM and to CP-violation in the electron-nucleus
interaction. Sophisticated atomic and molecular calculations are
required to account for these interactions.
%The observation of any finite EDM would be a very significant
%discovery; however, the interpretation of experimental results
%requires theoretical treatment on various levels. For instance, the
%most sensitive limits on the EDM of the electron come from
%experiments with paramagnetic atoms (e.g. $^{205}$Tl~\cite{Reg02})
%and molecules (e.g. YbF~\cite{ACME18}) and, therefore, the extraction
%of the fundamental fermion EDM relies on atomic and molecular
%calculations and must also account for possible other sources of CP
%violation, e.g. in the lepton-quark interaction.
Similarly, when using closed-shell atoms (e.g.\ $^{199}$Hg~\cite{Gra16}), atomic and
nuclear theory are both required to extract nucleon EDMs or even
quark EDMs and colour EDMs.\@ Somewhat easier is
the extraction of fundamental fermion EDMs from that of the
neutron~\cite{Abel20}, proton, or even light nuclei measurements.
Interestingly, today, of the fundamental fermions only the muon
provides an EDM limit from a direct measurement~\cite{Ben09}.

The known CP-violation of the electro-weak Standard Model~(SM) produces
EDMs only via higher-order loop contributions. 
These are at least five orders
of magnitude too small to be detected for current experimental
sensitivities.
However, most new physics scenarios include additional sources of
CP-violation which quite naturally could account for the observed
baryon asymmetry of the universe~\cite{Morrissey2012,Li2024}, and they typically predict much
larger EDMs: the experimental EDM bounds thus tightly constrain the
parameter space of such new-physics models and theories.

Hadronic EDMs, induced by the QCD $\theta$-term, serve also as direct probes of $\theta$, the last unmeasured SM parameter. The absence of observed hadronic EDMs constrains $\theta$ to an extremely small value ($\lesssim 10^{-10}$), defining the `Strong CP Problem.’ This also rules out naive Supersymmetry models, leading to the `SUSY CP Problem.’ 
Furthermore, the Peccei-Quinn mechanism~\cite{Peccei:1977hh} links $\theta$ to axions, suggesting a background axion field as a dark matter candidate. This would manifest as an oscillating nucleon EDM \cite{Abel2017PRX} at the axion frequency, making EDM searches a direct test for axions and their mass.

% For hadronic probes like neutrons, protons and nuclei, EDMs can also be induced by the so called $\Theta$-term of QCD. 
% Hadronic EDM measurements can thus also be considered as measurements of $\Theta$, which is the last unmeasured parameter of the SM. 
% The fact that nonzero hadronic EDMs were not yet observed constrains $\Theta$ to be extremely (perhaps unnaturally)
% small, of order $10^{-10}$: a situation termed the `Strong CP Problem'.
% Non-observation of any finite EDM so far also already
% excludes naive Supersymmetry models, and is known as the `SUSY
% CP Problem'. Additionally, the connection of $\Theta$ to axions via the Peccei-Quinn mechanism~\cite{Peccei:1977hh} led to the hypothesis of a background axion field to explain the Milky Way's dark-matter density. This would present as an oscillating nucleon EDM \cite{Abel2017PRX} with a frequency equal to the axion frequency, making searches for nucleon EDMS also direct probes of the axion and its mass.

The next round of planned experiments allows for a few orders of
magnitude improved sensitivity, further closing the gap to
the electro-weak SM EDMs\@.

\begin{table}[]
    \centering
    \begin{tabular}{ | c | c | c | p{7.5cm} |}
    \hline
    System & upper limit [$e$cm], 95\% C.L. & Reference & Comment \\ \hline
    n & $2.2\times10^{-26}$
%0.0\pm0.35\pm0.06$&$1.8\times10^{-26}$  90\% C.L.
    &\cite{Abel20} &direct limit  \\ \hline
    $\mu$ &$1.9\times 10^{-19}$&\cite{Ben09} &direct limit \\ \hline
    $^{199}$Hg &$7.4\times10^{-30}$  &\cite{Gra16} & best indirect (single-source) limit for proton: $|d_p| < 5.4\times 10^{-25}~e$cm, 95\% C.L.  \\ \hline
    %ThO &$-81.2\pm59.4\pm49.3 $ &$1.1\times10^{-29}$  90\% C.L.&\cite{ACME18} &used to set a limit for the electron  \\ \hline
    HfF$^+$ & $4.8\times 10^{-30}$ &\cite{Roussy23} &lowest EDM bound for any system; best indirect (single-source) limit for the electron: $|d_e| < 4.8\times 10^{-30}~e$cm, 95\% C.L.\\ \hline
    \end{tabular}
    \caption{The presently most precise EDM limits in each category: bare nucleon, bare lepton, closed-shell atoms and molecules,
    open-shell atoms and molecules. The extraction of, e.g., the electron or the proton EDM limits assumes a
    single source of CP violation, i.e., other EDMs or CPV interactions are neglected. For global analyses including multiple sources, see \cite{Degenkolb24,Gaul24,Chupp14}.}
        \label{tab:my_label}
\end{table}

% \begin{table}[]
%     \centering
%     \begin{tabular}{ | c | c | c | c | p{5.5cm} |}
%     \hline
%     System & $\Delta f $ ${ / \rm \mu Hz}$ & upper limit [$e$cm], 95\% C.L. & Reference & Comment \\ \hline
%     n & & $2.2\times10^{-26}$
% %0.0\pm0.35\pm0.06$&$1.8\times10^{-26}$  90\% C.L.
%     &\cite{Abel20} &direct limit  \\ \hline
%     $\mu$ & &$1.9\times 10^{-19}$&\cite{Ben09} &direct limit \\ \hline
%     $^{199}$Hg & &$7.4\times10^{-30}$  &\cite{Gra16} & best indirect (single-source) limit for proton: $|d_p| < 5.4\times 10^{-25}~e$cm, 95\% C.L.  \\ \hline
%     %ThO &$-81.2\pm59.4\pm49.3 $ &$1.1\times10^{-29}$  90\% C.L.&\cite{ACME18} &used to set a limit for the electron  \\ \hline
%     HfF$^+$ & $-7.3\pm11.4\pm3.5 $& $4.8\times 10^{-30}$ &\cite{Roussy23} &lowest EDM bound for any system; best indirect (single-source) limit for the electron: $|d_e| < 4.8\times 10^{-30}~e$cm, 95\% C.L.\\ \hline
%     \end{tabular}
%     \caption{The presently most precise EDM limits in each category: bare nucleon, bare lepton, closed-shell atoms and molecules,
%     open-shell atoms and molecules. The extraction of, e.g., the electron or the proton EDM limits assumes a
%     single source of CP violation, i.e., other EDMs or CPV interactions are neglected. For global analyses including multiple sources, see \cite{Degenkolb24,Gaul24,Chupp14}.}
%         \label{tab:my_label}
% \end{table}

\section{Theory of EDM and complementarity to high energy physics}
%Jordy and Roland
% one page presenting the european activities and international endeavors

Theory is essential for designing EDM experiments and interpreting results in terms of CPV sources. 
Computing EDMs, even within the SM, is challenging due to the non-perturbative nature of QCD and the many-body effects in complex systems like nucleons, atoms, and molecules. 
Beyond the SM, additional CPV operators further complicate the analysis. 
A robust theoretical framework is crucial to link EDM measurements to fundamental questions like the matter-antimatter asymmetry and new physics searches.

% Theory is necessary for preparation of EDM experiments and for interpretation of their outcome in terms of the underlying source(s) of CP violation at the level of particle physics.
% These are challenging problems. 
% For instance even within the SM of particle physics, computing EDMs of complex systems like nucleons, atoms, and molecules is not trivial. When considering beyond-the-Standard-Model physics the challenge is even greater because there are many more possible CPV operators at the level of elementary fields. Additional complications arise from the nonperturbative nature of QCD at low energies, making it difficult to connect CPV sources involving quarks and gluons (like the QCD $\Theta$-term or quark chromo-electric dipole moments) to the actual measurements. Another layer of complications arises from many-body nuclear, atomic, and molecular calculations that are necessary to prepare the experiments and subsequently to interpret the measurements of nuclei, atoms, and molecules. Finally, it is important to have a theoretical framework that makes it possible to connect EDM measurements to the broader search for beyond-the-Standard-Model physics and important outstanding questions such as the origin of the matter/antimatter asymmetry of the universe. 

Significant progress has been made on some of the above aspects that we briefly summarize here
\begin{itemize}
\item  General mechanisms of CP violation arising from high-energy particle physics have been described by the SM Effective Field Theory and evolved to lower energies using the renormalization group evolution \cite{Dekens:2013zca}. The operators involving quarks and gluons have been matched to hadronic CPV operators involving nucleons and mesons. Chiral effective field theory is then used to compute the CPV nuclear forces which are then used to compute EDMs and Schiff moments of atomic nuclei \cite{deVries:2020iea}. 
\item A major complication is the non-perturbative QCD matrix elements that connect hadronic/nuclear CPV to CPV at the quark/gluon level. Lattice QCD techniques are being used to compute these matrix elements from first principles. 
Several groups have attacked the neutron EDM arising from the QCD $\theta$ term finding, for the first time finding non-zero results \cite{Dragos:2019oxn,Liang:2023jfj} although the uncertainties are still large \cite{Liu:2024kqy}. Higher-dimensional operators are also being investigated but are more complicated. Interesting new methods are being investigated using the gradient flow \cite{Shindler:2021bcx}. 
\item First-principle calculations have appeared for the EDMs of light nuclei in terms of CP-odd hadronic interactions \cite{deVries:2011an,Froese:2021civ}. Following progress made in other symmetry-violating observables (such as neutrinoless double beta decay) there are attempts to perform ab initio calculations of Schiff moments of large nuclei. 
This is crucial to reduce the significant theoretical uncertainties that are currently plaguing the interpretation of diamagnetic atomic EDM experiments~\cite{Engel:2013lsa}.
\item  CPV operators involving heavy SM fields such as W-, Z-, and Higgs-bosons or top quarks can be probed with EDM experiments through quantum corrections \cite{Brod:2013cka,Ema:2022wxd}. 
Global analyses appeared that study the complementarity of EDM and collider experiments in probing, for example, CPV Higgs interactions \cite{Cirigliano:2019vfc}. These analyses show the great potential of EDM measurements as they provide, by far, the most accurate probes of CPV interactions, making EDMs indispensable in particle physics. For example, a future Higgs factory would measure the real parts of the Yukawa interactions between fermions and the Higgs boson while EDM experiments probe the imaginary parts. Both are necessary to fully test the Higgs sector of the Standard Model.
\item EDMs provide strong constraints on models of electroweak baryogenesis that involve new sources of CP violation. Certain minimal setups are already ruled out \cite{Huber:2006ri} and more elaborate theoretical model building is required to avoid the experimental limits. Such models are still viable, see e.g.~\cite{Athron:2025iew} and can be tested in next-generation EDM experiments. 
\item Electronic structure theory of atoms and molecules predicts internal electric fields $E_\mathrm{eff}$, crucial to relate a measured molecular EDM to the fundamental electron EDM (see below). Models and mean-field approaches that allow to compute these fields efficiently for a wealth of systems \cite{gaul:2019} have been complemented at the precision frontier by a host of systematically improvable many-body electronic structure methods to provide accurate predictions of $E_\mathrm{eff}$ and coarsely estimate their error bars \cite{haase:2021}.
\item Molecular theory has led to the identification of new candidates for eEDM searches such as RaF~\cite{isaev:2010,kudashov:2014}, for the study of which impressive experimental progress has been achieved at CERN~\cite{garciaruiz:2020,udrescu:2021,udrescu:2024}. Further candidate systems include BaOH or PaF$^{3+}$, with the latter being particularly sensitive to the Schiff moment of the ${}^{229}$Pa nucleus.
\item Theory has paved the way for novel experimental schemes such as laser-cooling of neutral diatomic~\cite{dirosa:2004} or polyatomic~\cite{isaev:2016} molecules. This technique features prominently in next-generation electron EDM experiments.
\item Theoretical approaches have been developed to predict sensitivities in atoms and molecules to all relevant CPV sources within these composite systems. These parameters have served as crucial ingredients for global analysis of atomic and molecular EDM searches \cite{Gaul24,Degenkolb24,Chupp14} and assist in selecting target systems that complement best ongoing searches for beyond-the-Standard-Model physics at high-energy and low-energy facilities and laboratories.
\end{itemize}

\section{The European EDM landscape and the global context}

\subsection{Searches for the neutron EDM}
%S. Degenkolb and D. Ries
% one page presenting the european projects and international competitors

The neutron EDM has a special role in the global context, both as the first EDM measurement to be performed historically and as the key experimental evidence underlying the Strong CP problem. It is also one of the systems in which experimental measurements most closely approach the expected finite signal from the Standard Model. Because interpreting the neutron EDM is to some extent more straightforward than for nuclear or atomic/molecular systems (which require additional input from hadronic and nuclear theory), and because the neutron EDM indeed also contributes to those systems, it offers a key opportunity to test the model-independent mechanisms by which CP-violation is expressed within the Standard Model. 

While early neutron EDM experiments were performed using polarized beams, modern experiments rely heavily on sources of ultracold neutrons (UCN). Facilities in Europe have played a leading role for many decades, for both UCN source development and neutron EDM measurements. All such experiments are statistically limited at present, with near-term improvements of up to an order-of-magnitude expected from currently available UCN sources in conjunction with external room-temperature spectrometers. The most recent and precise experimental bound \cite{Abel20} was measured at PSI, which now hosts the n2EDM experiment as its flagship at a solid deuterium UCN source. The PanEDM experiment based at the ILL exploits a newly available superfluid helium UCN source. Experiments based in North America at LANL and TRIUMF respectively exploit deuterium and helium sources, with similar measurement concepts and sensitivity goals. A two-cell differential spectrometer was first demonstrated by a PNPI collaboration, measuring at the ILL \cite{Serebrov15}. A particle physics beamline at the European Spallation Source could later host Beam EDM, a complementary approach using pulsed cold neutron beams rather than UCN.

Achieving sub-$10^{-27}e$cm sensitivity in future neutron EDM experiments will require new measurement techniques. 
In-situ experiments in superfluid helium~\cite{Golub1994,Ahmed2019} could enhance sensitivity by leveraging higher UCN densities and electric fields. 
Novel approaches are under investigation, driving an emerging international effort for next-generation neutron EDM measurements in Europe. 
Advances in UCN production and improved control of systematic errors now provide a foundation for a coordinated European research initiative.

\subsection{Searches for the electron EDM and CP violation in open-shell atoms, molecules}
%S. Hoekstra and M Trabutt
% one page presenting the european projects and international competitors

Measurements using open-shell atoms and molecules are sensitive to the electron EDM, $d_e$, and a CPV electron-nucleon interaction parameterized by $C_s$, though it is common to express results in terms of $d^{\rm equiv}_e$ alone. Taking this approach, the SM prediction is $d^{\rm equiv}_e = 1.0 \times 10^{-35}~e$cm~\cite{Ema:2022wxd} while for the bare electron $d_e = 5.4 \times 10^{-40}~e$cm~\cite{Yamaguchi2020}. Most theories beyond the SM predict values of $d_e$ many orders of magnitude larger~\cite{Chupp19}. 
Modern experiments exploit the enormous effective electric fields ($E_{\rm eff}$) of polar molecules, and try to make the number of molecules ($N$) and the spin precession time ($T$) as large as possible. Table \ref{tab:EDMProspects} lists planned and ongoing eEDM experiments, showing the species and method used and the anticipated sensitivity.

%One approach uses molecular beams and measures the change in the spin precession angle as the molecules fly through the experiment. This has the advantage of very large $N$, but offers only small $T$. The ACME collaboration (USA) has pioneered this approach using beams of metastable ThO molecules, setting the limit $|d_e| < 1.1 \times 10^{-29}~e$~cm~\cite{ACME18}. A third generation of this experiment is currently underway. Another approach, pioneered by the JILA team (USA), uses trapped molecular ions (HfF$^+$) and measures the spin precession frequency in the frame of a rotating electric field. This approach currently sets the best eEDM limit, $|d_e| < 0.41~e$~cm~\cite{Roussy23}. The trapped ion method gives very long $T$, but the density of ions has to be kept small so $N$ is much smaller with this approach. A third generation of this experiment uses a new species, ThF$^+$, which has a higher $E_{\rm eff}$ and offers even larger $T$.

\begin{table}
    \centering
\begin{tabular}{|c |c| c| c| c| c| c| } \toprule
 Collaboration & Species  & Method & Sensitivity & Status &  Duration  & Ref\\
 & & & ($10^{-29}~e$cm) &  & (years)  &   \\
 \hline
PanEDM I & n &  UCN & 380 & Commissioning & 5 & \cite{Wurm2019}\\
PanEDM II & n &  UCN & 79 & Commissioning & 8 & \cite{Wurm2019}\\
Beam EDM & n & beam & 500 & proof-of-principle & ? & \cite{Piegsa2013}\\
n2EDM & n & UCN & 110 & Start data-taking & 4 & \cite{Ayers2021}\\
n2EDMagic & n & UCN & 50 & Construction & 5 & \cite{Ayers2021}\\
nEDMsf & n& UCN & 20 & Development & 7 & \cite{Ahmed2019} \\
ACME III & ThO ($^3\Delta_1$) & Beam & 0.1  & Commissioning &  &\cite{Wu22} \\
JILA III & ThF$^+$ ($^3\Delta_1$) & Ion trap & & Commissioning &  & \cite{Ng22} \\
Imperial II & YbF ($^2\Sigma$) & $\mu$K beam & 0.1 & Commissioning & 3 & \cite{Fitch21} \\
Imperial III & YbF ($^2\Sigma$) & Lattice & 0.01  & Construction & 6 & \cite{Fitch21} \\
NL-eEDM I& BaF ($^2\Sigma$) & Slow beam & 0.5 & Commissioning & 3 & \cite{NLEDM} \\
NL-eEDM II& BaOH ($^2\Sigma$) & Lattice & 0.1 & Construction & 6 & \cite{NLEDM2} \\
PolyEDM & SrOH ($^2\Sigma$) & Lattice & & Construction & & \cite{Anderegg23} \\
EDM$^3$ & BaF ($^2\Sigma$) & Matrix & & Construction & & \cite{Vutha18} \\
DOCET & BaF ($^2\Sigma$) & Matrix &  & Construction & 3 & \cite{Messineo24} \\
EDMMA & Cs & Matrix &  & Construction &  & \cite{Battard23} \\
CeNTREX & $^{205}$TlF & Beam & &Commissioning &  & \cite{Grasdijk2021} \\
HeXe & $^{129}$Xe & $^3$He-comagnetometer & 10 & Construction & 4 & \cite{Allmendinger2019}\\
quMercury & $^{199}$Hg & ultracold atoms & 1 & Construction & 5 & \cite{Stellmer2024Trento} \\
ALADDIN & $\Lambda_c^+, \Xi_c^+$ & collider & $1\times10^{13}$ & Development & ?+2 & \cite{Akiba}\\
muEDM I & $\mu$ & frozen-spin & $4\times10^{8}$ & Commissioning & 3 &\cite{Adelmann2025} \\
muEDM II & $\mu$ & frozen-spin & $6\times10^{6}$ & Conception & 10 &\cite{Adelmann2025} \\
pEDM I & p & frozen-spin & 1 & Development & 5 & \cite{Alexander:2022rmq}\\
pEDM II & p & frozen-spin & 0.01 & Conception & 5 & \cite{Alexander:2022rmq} \\
\hline
\end{tabular}
\caption{Sensitivity goals of planned EDM measurements. All values are reported as transmitted in private communication or published in cited references.}
\label{tab:EDMProspects}
\end{table}

Several new approaches are being developed to reach uncertainties at the $10^{-31}~e$~cm level or better on a 5-10 year timescale. One promising avenue, pursued by the Imperial and NL-eEDM teams, uses trapped, neutral molecules cooled to \si{\micro\kelvin} temperatures so that $T$ can be large~\cite{Alauze21, Fitch21, NLEDM}. A second approach, adopted by PolyEDM and NL-eEDM uses polyatomic molecules which can be fully polarized in small electric fields and provide internal co-magnetometry~\cite{Anderegg23, NLEDM2}. A third method, followed by the EDM$^3$, DOCET, and EDMMA collaborations uses atoms or molecules embedded in a solid-state matrix resulting in enormous values of $N$~\cite{Vutha18, Messineo24, Battard23}.

\subsection{Searches for nuclear CP violation}

Due to the screening by the orbiting electrons, nuclear EDMs cannot be directly studied in neutral atoms and molecules placed in external electric fields. However, the interaction of nuclear CPV properties with electrons in atoms and molecules can still induce an overall atomic/molecular EDM\@. 
In closed-shell atoms and molecules, all electrons are paired, leading to a suppression of some leptonic contributions and enabling sensitive searches of nuclear CPV properties; the nuclear Schiff and magnetic quadrupole moments, as well as nuclear-spin-dependent CPV electron-nucleus interactions.

% A list of measurements and an experiment under commissioning using closed-shell systems to search for nuclear $CP$ violation is given in Table~\ref{tab:EDMProspects}. 
The most precise measurement with a closed-shell system to date used a vapor cell of $^{199}$Hg~\cite{Gra16}, achieving a very high number of probed atoms with $T=170$~s. 
Experiments with $^{225}$Ra~\cite{Bishof2016} and $^{171}$Yb~\cite{Zheng2022}  were performed in an optical dipole trap, while experiments with $^{129}$Xe were performed with SQUID detection of spin-polarized gases and $^{3}$He as a co-magnetometer~\cite{Sachdeva2019,Allmendinger2019}. 
Experiments with $^{205}$TlF employ rotationally cold beams, with the upcoming CeNTREX experiment also exploring laser cooling to use \si{\micro\kelvin} beams~\cite{Grasdijk2021} .

Several other experiments based on novel techniques are in different stages of development. Using noble gases, an experiment to measure the EDM of the radioactive $^{221,223}$Rn is in development at TRIUMF~\cite{Tardiff2014}, while an experiment with ultracold $^{199}$Hg prepared in a magneto-optical trap is being constructed at the University of Bonn~\cite{Lavigne2022,Stellmer2024Trento}. The design of a chip-based experiment using quantum logic spectroscopy with $^{171}$Yb$^+$ and $^{232}$ThF$^+$ ions is underway at the University of Nevada at Las Vegas~\cite{Zhou2024Trento}. 

Heavy species are strongly favorable for nuclear CPV searches, as sensitivity roughly scales with $Z^2$, and heavy and deformed nuclei have significantly enhanced CPV nuclear moments. As a result, experiments with such nuclei provide the highest sensitivity to symmetry violation in the fundamental forces. However, nuclei heavier than $^{208}$Pb are radioactive. Therefore, experiments with heavy radioactive molecules, while faced with significant technical challenges, are a promising avenue for future searches of nuclear CPV~\cite{ArrowsmithKron2024}. The FunSy experiment is currently under construction at CERN to study the electronic structure and assess the potential for nuclear CPV studies of radioactive molecules.

\subsection{Collider and storage ring searches for EDMs of charged particles}
%A. Keshavarzi and P. Schmidt-Wellenburg
% one page presenting the european projects and international competitors
A fixed target experiment at LHC, ALADDIN~\cite{Akiba}, proposes to measure the magnetic and electric dipole moment of the charm baryons, $\Lambda_C^+$ or $\Xi_c^+$ using two bent crystals to channel forward-produced charm baryon from the proton collision. This induces a spin precession which is measured in a magnetic spectrometer combined with at RICH detector. The method can also be extended to other hadrons and the $\tau$-lepton, providing highly sensitive probes to the Standard Model and to potential new physics which remained so far unexplored due to the challenges imposed by the short lifetimes of these particles.\\

Searches using storage rings which confine charged particles using electric, $E$, and magnetic fields, $B$, offer high-sensitivity probes of EDMs. 
The electric field in the radial direction toward the center of the storage orbit will induce a vertical tilt in the polarization of the stored particles if an EDM exists. 
%More specifically, as the electric field is perpendicular to the spin axis of the particle (and therefore perpendicular to the axis of the EDM), the particle's spin will precess in the vertical plane. 
Therefore, measuring the EDM requires storing highly spin-polarized particles and  
detecting a change in the polarization perpendicular to the electric field with time.
%with time is the signal for an EDM\@. 
Dedicated storage ring experiments exploit the frozen-spin technique~\cite{Farley2004}, either by
using only a radial $E$-field and a specific momentum or by adjusting the radial $E$-field in a magnetic storage ring such that $E\approx aBc\beta \gamma^2$ is fulfilled. Here, $a$ is the anomalous magnetic moment~(AMM), $c$ the speed of light, and $\beta$ and $\gamma$ the relativistic factors. 
A proposed search for the EDM of the bare electron uses yet another approach, a spin-transparent, figure-8 shaped storage ring with a prospected sensitivity of \SI{6e-30}{ecm}~\cite{Suleiman2023PLB}.

The first option is used for particles with a large AMM (e.g.\ proton) thereby ``freezing'' the dominant magnetic spin-precession component to the orientation of the momentum and leaving it precess purely by the EDM component (to first order) coupling to the magnitude of the $E$-field.
In the second option for small AMM (e.g.\ muon) the spin is also ``frozen'' to the momentum of the particle and the spin precesses around the $E$-field in the rest frame of the particle $\gamma c |\vec{\beta}\times\vec{B}| \approx \mathcal{O}(\SI{1}{GV/m})$ for $B=\SI{3}{T}$ and $\beta\approx0.75$ for the muon.
 % With the AMM component nullified and the polarization of the stored particles at injection known, the appearance of a vertical polarization component of the particle's spin with time is the signal for a non-vanishing EDM\@. 
~\\

The search for a muon EDM, being mounted at the Paul Scherrer Insitute~\cite{Adelmann2025}, is the first deploying the frozen-spin in a compact storage trap. 
In a first phase the collaboration aims to demonstrate the frozen-spin method and search for a muon EDM using an existing solenoid with a sensitivity of \SI{4e-21}{ecm}, sufficient to improve the current best measurement~\cite{Ben09} by more than two orders of magnitude. In the second phase, after 2030, an instrument featuring a dedicated trap solenoid will be used to profit from the highest-intensity muon beam at PSI, resulting in a sensitivity of better than \SI{6e-23}{ecm}.
By modifying the muEDM experiment it is possible to search for CPV in light beta-decaying isotopes, inspired by Ref.~\cite{Khriplovich}, using partially stripped ions at ISOLDE\@.\\

% \vspace{3.5mm}
% \paragraph*{The Proton EDM Experiment}~\\
% % {\color{red} AK: 
The Proton EDM~(pEDM) experiment~\cite{Alexander:2022rmq,CPEDM:2019nwp} is an international initiative for a first direct search for the proton EDM and will also employ the frozen-spin technique to achieve a Phase-1 sensitivity of $10^{-29} e\cdot{\rm cm}$. This will improve the current (indirect) limit by $\mathcal{O}(10^4)$, being sensitive to BSM mass scales ranging from $\mathcal{O}$(1~GeV) $\rightarrow$ $\mathcal{O}$(1~PeV). 
% pEDM will also be sensitive to the axion-induced oscillating proton EDM for axion frequencies/masses ranging from 1~mHz $\rightarrow$ 1MHz/$10^{-22}$~eV  $\rightarrow$ $10^{-7}$~eV, and will 
Significant prototyping of the experiment was completed at the COSY ring in Jülich (Germany)~\cite{CPEDM:2019nwp} and is ongoing at BNL and in the UK\@.
Major design, engineering, and simulation work has already been completed for two possible locations: Brookhaven National Laboratory~(BNL) and 
CERN~\cite{pEDM2,pEDM3,pEDM4,pEDM5,pEDM6,pEDM7,pEDM8,pEDM9,pEDM10,pEDM11,pEDM12,pEDM13,pEDM14,pEDM15,pEDM16}.
% The experimental design, engineering and modeling are at an advanced stage. The measurement techniques are well understood, and the key systematic effects have been simulated and quantified. 
A similar approach is proposed for deuterons and fully stripped $^{3}$He$^{2+}$ by the JEDI and srEDM collaborations~\cite{Rathmann2013}. 
R\&D for pEDM's Phase-2 Experiment is underway with the specific aim to be sensitive to the SM prediction of $10^{-31} e\cdot{\rm cm}$.
%

%%%%%%%%%%%%%%%%%%%%%%%%

\section{Challenges and opportunities of smaller scale precision experiments}
There are a number of issues pertaining to smaller-scale precision experiments competing for funding with the extremely large projects that are common in the field of particle physics.
These include:
\begin{itemize}
\item{} The EDM community, though small and widely dispersed, is actively strengthening communication through workshops and schools, fostering collaboration and knowledge exchange.

\item{} These experiments thrive on the dedication of a small, highly skilled team with diverse expertise. Ensuring robust backup plans and cross-training enhances resilience, enabling steady progress even in the face of challenges like recruitment gaps, illness, or other unforeseen circumstances.

% These experiments are often carried out by a very small number of people across a wide range of expertise making them extremely dependent upon critical individuals.  
% A single person's illness, or difficulty of recruitment in a specific area, can impede progress significantly if adequate backup plans are not in place.

\item{} The breadth of physics involved often makes recruitment challenging.
%In more conventional particle physics,
%there are many candidates available with expertise in the necessary techniques.
Finding high-quality candidates with expertise in, e.g., particle physics,
spin physics and polarimetry, modeling and simulation,
magnetometry, surface physics, atomic laser spectroscopy, vacuum techniques, cryogenics, and high-voltage techniques who wish to work on a
particle-physics related experiment is not always straightforward.

\item{} 
In several cases, CERN provides extensive resources and technical support for experiments conducted on-site, while facilities like ILL and the future ESS operate as user-driven platforms, offering neutrons but with less direct technical assistance. 
Strengthening collaborations between host institutions and researchers in precision experiments can enhance support, optimize resource use, and reduce the need for frequent equipment transfers, fostering more efficient and effective research.

% A laboratory like CERN often becomes intimately involved in the
% experiments carried out there, and provides resources and technical
% support. 
% Other institutions such as the ILL or the future ESS are regarded as ``user
% facilities:'' the institution's responsibility ends where neutrons
% emerge from a beamline, and considerably less technical support can
% be provided for users. 
% By promoting active collaborations between hosting institutes and users for precision experiments one might overcome these particular challenges, and reduces necessary shipping of equipment to and from sites elsewhere where the actual development takes place.

\item{} In large projects like LHC collider experiments at CERN, technology is well-established before operations begin, often following extensive R\&D\@. Once the design is finalized, detailed modeling allows for precise rate predictions. In contrast, high-precision experiments achieve their ultimate sensitivity through continuous refinements and incremental improvements over several years, highlighting the importance of adaptability and ongoing optimization in pushing the frontiers of discovery.

\item{} The drive for continuous scientific output is strong. 
In large multi-purpose experiments, multiple measurements occur simultaneously, leading to hundreds of publications annually. 
In contrast, precision experiments focus on measuring a single parameter with exceptional accuracy, often requiring years to achieve a groundbreaking result, underscoring their unique contribution to advancing fundamental science.
\end{itemize}

\subsection{National laboratories and university labs}
Although EDM experiments are small in comparison with more
conventional particle physics projects, they can be relatively large
when compared with other areas of physics. For neutrons, muons, and
highly charged ions, their scale typically makes it already impossible for a single
institution to carry out the work. Substantial support, engineering
assistance and so on from national and international laboratories is therefore vital.
Universities provide academic staff, students, and research associates to
carry out underpinning research and development.

\subsection{Education and outreach}
EDM experiments offer an unparalleled training ground for students, immersing them in every stage of experimental physics — from conceptualization and modeling to design, construction, measurement, analysis, and publication. This hands-on experience cultivates versatile and highly skilled experimentalists.

While they may not have the high-profile imagery of large-scale particle physics projects, EDM experiments provide an excellent opportunity for long-lasting outreach. 
By showcasing the elegance of precision science, they can inspire future generations to pursue STEM fields and effectively communicate the value of fundamental research to the public.

% EDM experiments do not have the high profile of their conventional
% particle-physics partners. The science of CP violation is
% challenging to explain to a lay audience, and the associated imagery
% is often not as dramatic.  Nonetheless, there is considerable scope
% to build and develop a substantial outreach program to inspire
% pupils to study science and to justify to the general public the
% expenditure of money on such projects.  The nature of the
% experiments also provides substantial opportunity to educate
% students in a wide range of physics-related skills, from
% computational modeling and analysis through to high-voltage,
% surface physics and vacuum technology. It must also be emphasized
% that the scale of the experiments allows very good students to learn
% a complete set of skills for an experimentalist covering all steps
% from ideas via modeling, design and construction to building
% equipment, measuring, analyzing and publishing. It is vital that
% particle physics maintains experimental activities on that level in
%order to contribute to the education of 'complete' experimentalists.

\subsection{Facilities}

Precision particle physics, including EDM research, relies on maintaining and advancing experimental facilities and particle sources with high quality, intensity, and brightness.
Europe hosts several world-leading facilities essential to this mission, such as ILL, PNPI, FRM-2, PSI and in the future ESS for neutron and ultracold neutron physics; ISOLDE at CERN for radioactive ions; and PSI for high-intensity muons.
Lattice calculations for precision QCD predictions profit from easy access to the newest and most powerful computing infrastructure. Likewise, large-scale electronic structure theory calculations require availability of and access to high-performance compute facilities and supercomputer centers.  
% Neighboring subfields benefit from unique opportunities at GSI and CERN’s AD/ELENA. 
Sustained support for these facilities and the university groups collaborating on their projects is crucial to preserving a diverse and robust particle physics community.

\section{Role of searches for EDM in the European strategy for particle physics}

\begin{enumerate}

\item{}The wide arena of particle physics,
including both the high-energy collider frontier and the low-energy precision-measurement frontier,
brings important new knowledge about our origins and about the nature of the Universe,
and stimulates the development of new technologies.
{\em Europe should maintain and strengthen its central position in particle physics, which is founded
on its strong national and international institutes, universities and laboratories.}

\item{} Already today, EDM experiments are a unique way of testing CPV in BSM at sensitivities at energy scales beyond any future colliders, above 100\,TeV assuming a large CPV phase.
Together with results of other high precision experiments at energies available today they may provide valuable guidance to the high-energy phsyics~(HEP)community in decisions on future accelerators and instrumentation. 
{\em Europe should strengthen funding of high precision experiments in national laboratories and universities for guidance and in theory and design of high energy particle physics facilities.}

\item{} EDM experiments require expertise over many areas of physics: particle physics,
spin physics and polarimetry, modeling and simulation,
magnetometry, surface physics, atomic laser spectroscopy,
high-voltage techniques, vacuum techniques, cryogenics and so on.
{\em EDM measurements provide a valuable route for education and
skills training of upcoming generations of physicists, and
studentships in this area should be supported.}

\item{} There are many different EDM projects, and these are themselves a subset of the many ongoing non-accelerator projects throughout Europe and around the world.  
{\em The EDM community needs to work with national and international particle and
astroparticle physics organizations to contribute to a coordinated
strategy.}

\item{} Neutron sources are primarily used for scattering and other applied-physics measurements.
{\em There is nonetheless a need to develop and maintain neutron sources appropriate for fundamental-particle studies.}

% \item{} In more general terms, high intensity, low momentum particle
% beams of ions, muons, and hadrons are required for
% various important precision experiments. {\em The strategy should
% include the need to develop and maintain these facilities and to
% make maximal use of the complementarity of the various research
% projects.}

\item{} EDM results are highly valued by theoretical physicists, as reflected in their strong citation impact. 
However, there is great potential to strengthen collaboration between the EDM experimental and HEP theory communities.
Enhancing this exchange -- akin to the synergies seen in broader particle physics -- could lead to deeper insights, innovative approaches, and a more integrated effort in exploring fundamental CPV phenomena. 
{\em Strengthening collaboration between experimentalists and theorists is essential for advancing EDM research. Likewise, fostering closer ties between HEP physics, low-energy precision physics, and related fields such as nuclear, atomic, and molecular physics will enable a more unified and interdisciplinary approach, driving deeper insights and innovation.}

\item{} Small-scale, elegant high-precision measurements provide an ideal vehicle to inspire students and the public, and to promote a general interest in fundamental physics. 
{\em There is a need to develop and implement a coherent outreach strategy.}

\item{} Techniques and methods used in EDM measurements are also used in medical applications and high precision metrology such as NMR (MRI) and atomic clocks. 
{\em Technology transfer from the field should continue to be promoted.}

\end{enumerate}

\end{document}